# ARTICLE

# Stoichiometric $Bi_2Se_3$ Topological Insulator Ultra-Thin Films Obtained Through a New Fabrication Process for Optoelectronic Applications

Matteo Salvato,*[a,b] Mattia Scagliotti,[a,b] Maurizio De Crescenzi,[a,b] Paola Castrucci,[a,b] Fabio De Matteis,[c] Michele Crivellari,[d] Stefano Pelli Cresi,[e] Daniele Catone,[e] Thilo Bauch[f] and Floriana Lombardi[f]

A new fabrication process is developed for growing $Bi_2Se_3$ topological insulators in the form of nanowires/nanobelts and ultra-thin films. It consists of two consecutive procedures: first $Bi_2Se_3$ nanowires/nanobelts are deposited by standard catalyst free vapour-solid deposition on different substrates positioned inside a quartz tube. Then, the $Bi_2Se_3$, stuck on the inner surface of the quartz tube, is re-evaporated and deposited in the form of ultra-thin films on new substrates at temperature below 100 °C, which is of relevance for flexible electronic applications. The method is new, quick, very inexpensive, easy to control and allows obtaining films with different thickness down to one quintuple layer (QL) during the same procedure. The composition and the crystal structure of both the nanowires/nanobelts and the thin films is analysed by different optical, electronic and structural techniques. For the films, scanning tunnelling spectroscopy shows that the Fermi level is positioned in the middle of the energy bandgap as a consequence of the achieved correct stoichiometry. Ultra-thin films, with thickness in the range 1-10 QLs deposited on n-doped Si substrates, show good rectified properties suitable for their use as photodetectors in the ultra violet-visible-near infrared wavelength range.

## 1. Introduction

In the last decade, Topological Insulators (TIs) have emerged as a new state of matter with the unique physical property to be insulators in the bulk and metallic on the surface.[1] Here the two dimensional (2D) surface electrons are characterized by Dirac dispersion and spin-momentum locking property, which prevents backscattering, if time reversal symmetry is preserved.[2] Compared to other 2D materials, as for example Graphene, TIs have the advantage of spin momentum locking property and of a finite bulk bandgap both offering new opportunities in fields such as spintronics, superconducting quantum computing,[3] and optoelectronics.[4] From this perspective, $Bi_2Se_3$, with its high surface mobility[5] $\mu$=10$^3$-10$^4$ cm$^2 \cdot $V$^{-1} \cdot$s$^{-1}$ and large bulk energy bandgap $E_G$=0.32 eV,[6] is the most promising among the TIs to explore broadband photo detection ranging from ultraviolet (UV) to infrared (IR)[7] and THz.[8]

TIs have a crystal structure ruled by van der Waals interactions, which is suitable for their integration with various other materials and, in particular, with Si.[9] The crystal structure of $Bi_2Se_3$ is rhombohedral with a unit cell that can be schematically represented by three blocks stacked along the [001] crystallographic direction, each formed by five atomic layers in the sequence Se-Bi-Se-Bi-Se, with lattice parameters $a=b$=0.416 nm and $c$=2.861 nm.[10] The layers inside each block are strongly ionic bounded whereas the different blocks, also known as quintuple layers (QLs), are weakly bounded together by van der Waals forces. The weak van der Waals bonding confers to $Bi_2Se_3$, and in general to all TIs, the interesting property to be grown in units of QLs and, in the case of thin films, on almost any kind of substrate independent on the lattice mismatch.[9,11] These characteristics, namely a wideband optical absorber with a metallic surface, together with the easiness to be integrated with other semiconducting materials, has motivated the recent use of $Bi_2Se_3$ as a component for optoelectronic nanodevices.[12] In particular, the possibility to operate from UV to IR is of paramount to detect signals from scintillators, for their use in high-energy astrophysics, as well as for operating in the data communication optical wavelength band.[4] From this point of view, the control of the thickness, favoured by van der Waals epitaxy, allows to achieve heterojunctions with widely tunable bandgap depending on the TI thickness.[7]

TI thin films are obtained with different techniques such as molecular beam epitaxy (MBE),[11] pulsed laser deposition (PLD),[13] chemical vapour deposition (CVD),[14] or vapour solid deposition (VSD)[5,15] to mention a few. Despite the great

[a.] Dipartimento di Fisica, Università di Roma "Tor Vergata", 00133 Roma, Italy.
[b.] INFN, Roma "Tor Vergata", 00133 Roma, Italy
[c.] Dipartimento di Ingegneria Industriale, INSTM and CiMER, Università di Roma "Tor Vergata", 00133 Roma, Italy.
[d.] Micro-Nano Characterization Facility, Fondazione Bruno Kessler (FBK), 38123 Trento, Italy.
[e.] CNR-ISM, Division of Ultrafast Processes in Materials (FLASHit) Area della Ricerca di Roma "Tor Vergata", 00133 Roma, Italy.
[f.] Quantum Device Physics Laboratory, Department of Microtechnology and Nanoscience, Chalmers University of Technology, 41296 Goteborg, Sweden.
*Corresponding author: matteo.salvato@roma2.infn.it

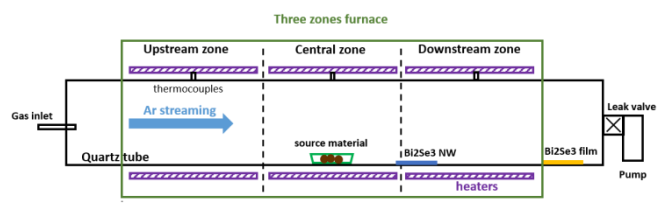

**Fig. 1** Sketch of the furnace showing the three heated zones, the quartz tube, the position of the source material and the substrates during the two procedures of growth. The big green rectangle represents the furnace walls.

versatility of TI materials to be grown with different techniques and on almost any kind of substrate, the experimental results reported in literature show $Bi_2Se_3$ films and nanoribbons lacking in Se atoms. This gives an n-doped character to $Bi_2Se_3$ with a sizeable shift of the Fermi level crossing the conduction band,[16,17] which tends to obscure the topological properties of the material. In this paper, we report a new fabrication method based on VSD technique where stoichiometric $Bi_2Se_3$ ultra-thin films are obtained using the by-products of a previous nanowire/nanobelt growth process. The quasi-perfect stoichiometry of the thin films has been assessed by x-ray photoelectron spectroscopy (XPS) and confirmed by scanning tunnelling spectroscopy (STS), revealing the energy position of the Fermi level in the middle of the bulk energy gap crossed by metallic Dirac states. The proposed process is very flexible, and it gives the opportunity to obtain different devices during the same fabrication process in a very inexpensive reproducible way and to open new routes towards large-scale production of TI based nanodevices.

## 2. Synthesis and characterization of the samples

The deposition system consists of an MTI Corporation three-zone furnace suitable for CVD and VSD processes. The furnace is equipped with a quartz tube positioned along the three zones. Each zone temperature is separately programmable and controlled by K-type thermocouples placed in the middle of each zone in close contact with the quartz tube surface. For a sketch of the deposition system, refer to Fig. 1. Depending on the process, an appropriate gas can be injected in the quartz tube from the upstream end and pumped by a mechanical pump positioned at the end of the downstream zone. The fabrication process consists of two consecutive procedures. During the first procedure, $Bi_2Se_3$ nanowires/nanobelts are deposited on different substrates, which are positioned across the boundary between the central zone and the downstream zone. After removing these first samples, new substrates for thin films deposition are loaded inside the quartz tube at the far end low temperature downstream zone, which is outside the heated regions of the furnace (see Fig. 1). With the tube under vacuum, the second procedure of the process starts: the temperature of all the three furnace zones is raised to re-evaporate the species stuck around the inner surface of the quartz tube during the first procedure. The re-evaporated species flow towards the pumping system depositing around the coldest walls of the tube where the new substrates are positioned. The final products are nanowires/nanobelts (first

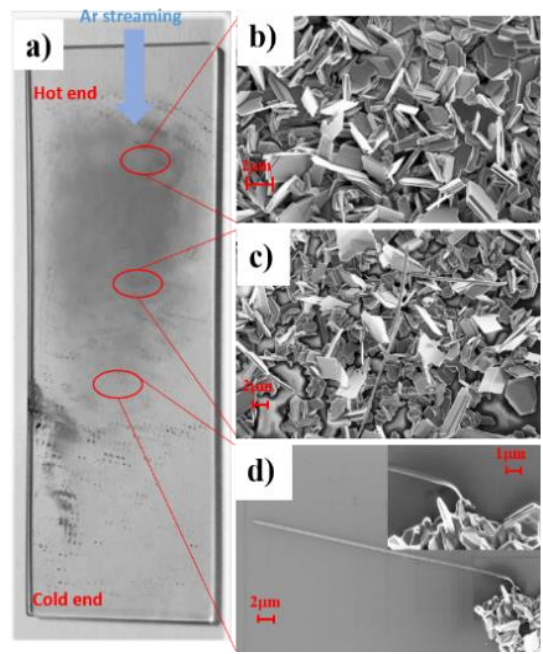

**Fig. 2** a) Optical image of a 75 mm long glass substrate after the first procedure. The species are scattered along the longitudinal length. b) SEM image acquired in the hottest zone of the substrate showing only the presence of nanoplatelets; c) nanowires start to grow at a moderate temperature in the middle of the substrate; d) nanobelt emerging from a platelet. Inset: particular of nanobelt growth from a nanoplatelet edge.

procedure) and ultra-thin films of different thickness deposited on different substrates (second procedure).

### 2.1 First procedure: deposition of nanowires/nanobelts

This part of the process and the mechanism of growth of nanowires/nanobelts is well documented in ref. [5]. Briefly, the source material $Bi_2Se_3$ powder is placed in a quartz boat positioned inside the quartz tube in the middle of the central zone of the furnace (see Fig. 1). The substrates are positioned at the same distance from the source and in correspondence of the boundary between the central zone and the downstream zone of the furnace where a temperature gradient can be generated by heating the two zones at different temperature. With the quartz tube sealed at the pressure of $1\times10^{-1}$ mbar, the temperature of the central zone and the downstream zone is ramped during 45 minutes from room temperature up to 585 °C and 340 °C respectively and left at these temperatures for 15 minutes. This allows the evaporation of the whole $Bi_2Se_3$ source material loaded in the quartz boat. The furnace heaters are then turned off and the temperature naturally drops allowing the evaporated species to condense on the substrates and all around the inner surface of the tube, mainly in its coldest zone close to the source material, where the substrates are located. $Bi_2Se_3$ nanoplatelets are formed at this stage and scattered all along the substrate surface with their sizes independent on the kind of substrate. When the temperature in the central zone reaches 540 °C, Ar flow is introduced inside the tube at a rate of 5 sccm maintaining the total pressure at 35 mbar. Fig. 2a shows an optical image of a 25x75 mm² glass substrate when the nanowire/nanobelt growth is over. The density of the species changes along the longitudinal direction, which corresponds to the Ar stream direction. Scanning electron microscopy (SEM) images in the same figure report the sample morphology in different zones of the substrate showing nanoplatelets, nanowires and nanobelts. The SEM analysis suggests that the growth of nanowires/nanobelts depends on the substrate temperature. In fact, they are absent in the hot end side of the substrate where only nanoplatelets are present (Fig. 2b). Going towards the coldest zone, nanowires/nanobelts start to appear and their density decreases with the density of nanoplatelets (Fig. 2c and 2d). **The composition of the deposited species is monitored, after the growth, by Energy Dispersive Spectroscopy (EDS). The analysis, performed on areas as large as 10 μm x 10 μm, gives an average composition of Se 56±2 % and Bi 44±2 %. This indicates a lack of Se in the average composition. The local EDS analysis, performed on single nanowire/nanobelt, gives the more correct result of Se 59±2 % and Bi 41±2 % indicating that the nanostructures grow in form of nanowires/nanobelts only when the correct atomic composition is achieved.**[5] Fig. 2d suggests the mechanism of growth of nanowires/nanobelts, which start from a nanoplatelet that acts as a precursor.[5] This is clearly observed in the inset of Fig. 2d where a nanobelt emerges from the edge of a nanoplatelet. Nanowires/nanobelts as long as tens of microns and thickness in the range 10-100 nm are routinely obtained with this method on all the mentioned substrates. The same morphology of the species shown in Fig. 2 is observed on all the loaded substrates and, in particular, on quartz substrates. This suggests that nanoplatelets as well as nanowires/nanobelts are also deposited all around the inner wall of the quartz tube.

**2.2 Second procedure: growth of ultra-thin films**

When the temperature of the tube drops down to 100 °C, the substrates covered with nanowires/nanobelts are unloaded and new substrates for thin film deposition are loaded in the quartz tube. These are Si, Si/SiO$_2$, glass and quartz. For optoelectronic device applications, pre-patterned substrates are also used. They consist of n-doped Si(00*l*) wafers with the top surface partially covered with a Pt electrode thermally deposited on SiO$_2$ template layer (for a full description of these substrates, hereafter named Si-Pt, see below). Occasionally, kapton tape was stuck on a glass substrate for testing the method as suitable for flexible electronic production. The substrates are positioned in the far downstream region of the quartz tube which is not directly heated by the furnace heaters (see the sketch in Fig. 1 and the picture in Fig. 3a). After flushing with Ar gas and pumping the tube several times, the temperature in all the three zones is raised up to 625°C in 20 minutes and left at this temperature for 30 minutes. This promotes the complete outgassing of the species attached to the inner wall of the tube during the first procedure of the process. The substrate temperature, measured by a power meter, never exceeds 100 °C during this procedure. This is of great importance because it allows using low temperature melting materials (e.g. flexible polymers) as well as multilayers and metallic prepatterned

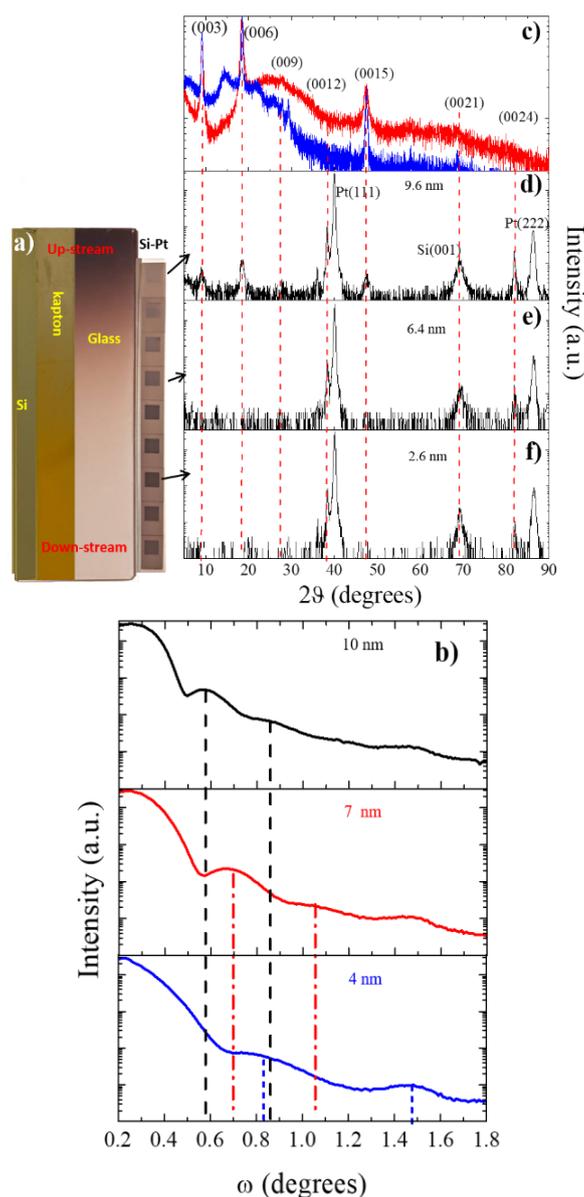

**Fig. 3** a) Optical image of Si, plastic kapton tape, glass, and Si-Pt substrates positioned as loaded for the film growth. The thickness of the Bi$_2$Se$_3$ films, deposited on Si-Pt substrates, are, from the top, 9.6, 8.6, 7.5, 6.4, 5.9, 3.8, 2.6, 1.3, 0.6 nm. b) Typical x-ray reflectivity measurements acquired on three different Bi$_2$Se$_3$ films deposited on 5 mm long Si substrates. The result of the thickness measurement is reported in each panel. c) XRD of the Bi$_2$Se$_3$ film deposited on glass (red) and plastic kapton tape (blue). d)-f) XRD for films deposited on Si-Pt substrates indicated by the arrows. The thickness of the Bi$_2$Se$_3$ films are reported in each panel.

wafers as substrates since the interdiffusion is prevented at this low temperature. During the heating up and dwelling time, the tube is pumped down to 10$^{-2}$ mbar. The outgassed Bi and Se species flow towards the downstream region where they meet a lower temperature zone in the part of the tube where the substrates are positioned (see Fig. 1). The strong temperature gradient promotes the Bi$_2$Se$_3$ condensation on the tube wall and on the substrate surface. To passivate the substrate surface by the possible presence of dangling bonds (especially for SiO$_2$), few mgs of pure Se are loaded in the central zone of the furnace.[11,18] **Since the Se evaporation temperature is around** 230 °C, it evaporates during the temperature ramp and deposits in part on the substrates and in part on the tube walls adding up to the species that are stuck on the tube walls before their re-evaporation takes place. The re-evaporation starts at higher temperature producing $Bi_2Se_3$ vapour with a corrected stoichiometry due to the extra Se added to the Bi and Se species already present on the tube walls. Fig. 3a shows an optical image of the film deposited on a glass, Si, kapton tape and Si-Pt substrates. The glass substrate shows a gradient in its transparency, which corresponds to a gradient in the film thickness, resulting in a thinner film in the downstream region.

**2.2.1 Ultra-thin film analysis and characterization.**

One of the critical points of this deposition method, and more generally of VSD, concerns the thickness of the samples. The species are deposited on the substrates under the action of the pump streaming, which leads to a gradient in the thickness along the longitudinal position. Temperature gradient effects can be excluded because of its reduced value (<100°C) in this zone. To evaluate properly this thickness gradient, we have performed x-ray reflectivity measurements every 5 mm of the Si substrate length. Fig. 3b shows the typical Kiessig fringes obtained from the measurements on three different positions, which allowed calculating a thickness gradient of 0.14±0.02 nm·mm$^{-1}$.[19] X-ray reflectivity is also used as a post deposition method for film overall thickness control. Based on the film thickness measurement, the amount of $Bi_2Se_3$ loaded in the quartz boat at the beginning of the first procedure is properly changed in order to obtain the desired film thickness.

The crystal structure of the films has been investigated by XRD measurements in grazing incident geometry to reduce the effect of the substrates on the x-ray count rate. The experimental spectra are reported in Figs. 3c-3f. The red and the blue spectra in Fig. 3c refers to the film deposited on the glass and kapton, respectively. Despite the amorphous substrates, the film grows well oriented along the [001] crystallographic direction. Moreover, the presence of the reflections identified as the (00*l*) orientations of the $Bi_2Se_3$ phase confirms the success of the method as promising in depositing films on flexible membranes, a paramount result in view of new applications of this material. Figs. 3d-3f show XRD measurements performed on $Bi_2Se_3$ films with different thickness deposited on Si-Pt substrates. As for the glass and kapton tape substrates, the samples grow (00*l*) oriented starting from the first atomic layers as shown in Fig. 3f where the XRD data of a sample with a thickness of 2.6 nm, roughly corresponding to 3 QLs, are reported. Moreover, no shift in the peak position is observed with increasing the film thickness, indicating that no strain is induced by the substrate on the $Bi_2Se_3$ lattice parameters. The same lattice orientation achieved on different substrates confirms the van der Waals growth mechanism for these films.

Despite the change in thickness, the composition remains constant along the entire 75 mm long substrates as confirmed by XPS analysis shown in Fig.4. **This was performed by using a non-monochromatic Mg Kα radiation (1253.6 eV) and a Riber**

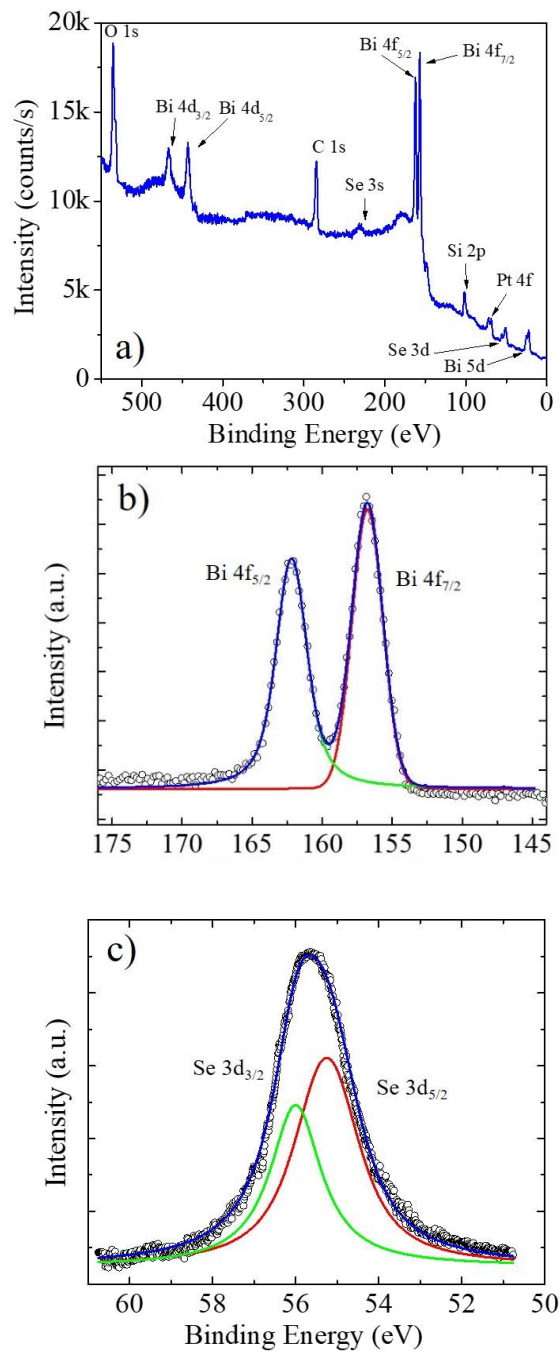

**Fig. 4** a) Extended XPS spectrum of the 9.6 nm thick $Bi_2Se_3$ film deposited on Si-Pt substrate.  b) $4f_{7/2}$ and $4f_{5/2}$ emission lines of Bi and c) $3d_{5/2}$ and $3d_{3/2}$ emission lines for Se. The open circles are the experimental data; the lines are fits to the data.

**MAC2 electron analyzer operating in retarding mode. In these conditions the total resolution of the core spectra reported in Fig.4 was estimated to be 0.8 eV. The energy scale was calibrated with reference to the binding energy of the C 1s peak measured to be 284.6 ± 0.8 eV with respect to the Fermi level.** Fig. 4a shows an extended XPS spectrum obtained for a 3.8 nm thick $Bi_2Se_3$ film deposited on the Si-Pt substrate. The presence of carbon and oxygen peaks comes most likely from air exposure contamination. **Apart from the Bi and Se peaks,**

and the signal coming from the Si-Pt substrate, only C and O contaminants due to the air exposure after the sample synthesis were detected. More extended measurements up to 1100 eV did not show any presence of other possible contaminants. For a quantitative analysis, the Bi $4f_{7/2}$ and $4f_{5/2}$ (Fig. 4b) and the Se $3d_{5/2}$ - $3d_{3/2}$ (Fig. 4c) core level peaks are considered using the Tougaard's method[20] for background removal and taking the final height of the curves as fixed parameters. Fitting the peaks with a Gaussian model using a Levemberg Marquardt algorithm, the atomic fraction $C_x$ of Bi and Se is calculated, which gives $C_{Bi}$=0.39±0.01 and $C_{Se}$=0.61±0.01. The same measurements performed on the 9.6 nm thick $Bi_2Se_3$ film give $C_{Bi}$=0.40±0.01 and $C_{Se}$=0.60±0.01 confirming that the $Bi_2Se_3$ composition is homogeneous, within the experimental uncertainty, along the entire substrate length. We emphasize that all the peaks show only one component excluding bonds with oxygen or carbon. In particular, no peaks between 159 eV and 160 eV, expected for $Bi_2O_3$, and between 58 eV and 60 eV, expected for $SeO_2$, are observed in Fig. 4b and Fig. 4c respectively.[21] Moreover, since XPS is a surface sensitive analysis, these results indicate that the Se passivating layer, eventually deposited on the substrate before the film growth, does not affect the correct stoichiometry which is achieved along the $Bi_2Se_3$ growth direction starting from the first deposited QLs.

The surface morphology of the films deposited on the Si and on the Pt regions of the Si-Pt substrates (see picture in Fig. 5a) is investigated by scanning tunnelling microscopy (STM) at room temperature in ultra-high vacuum (UHV) conditions and without any treatment of the sample surface. **STM imaging was performed using an Omicron-STM system with electrochemically etched tungsten tips. To remove the oxide layer from the tip and to ensure a sharpened shape, a high current of 50 nA and a 10V voltage was applied in tunnelling conditions. The STM atomic spatial resolution and calibration was performed by acquiring atomically resolved images of a bare HOPG, freshly peeled before the insertion into the UHV apparatus.[22] All images were acquired in the constant current mode and were unfiltered apart from the rigid plane subtraction. Typical voltage and current acquisition parameters were $V$=0.5V and $I$=0.5 nA respectively.** Fig. 5b shows an STM image of an extended area of a 9.6 nm thick $Bi_2Se_3$ film deposited on the Si side. The surface appears quite flat at nanoscale level with the presence of flat and striped regions. The stripes are evidenced in figure 5c where the profile acquired along the white line shows a periodicity of 2.8 ± 0.1 nm as reported in the top inset of the same figure. The origin of the stripes on TIs has been debated by many authors[23,24] and we will discuss it elsewhere.[25] They are commonly related to the substrate lattice mismatch[24] or to an intrinsic strain of the top $Bi_2Se_3$ unit cell.[23] In all cases, they are ascribed to a surface charge modulation rather than to a structural surface corrugation. **Fig. 5d shows a grain boundary between two regions where the stripes are not detected. The profile along the blue straight line is reported in Fig.5f where the decrease of the signal corresponds to the grain separation. A surface roughness of 0.1 nm is estimated on both sides of the grain**

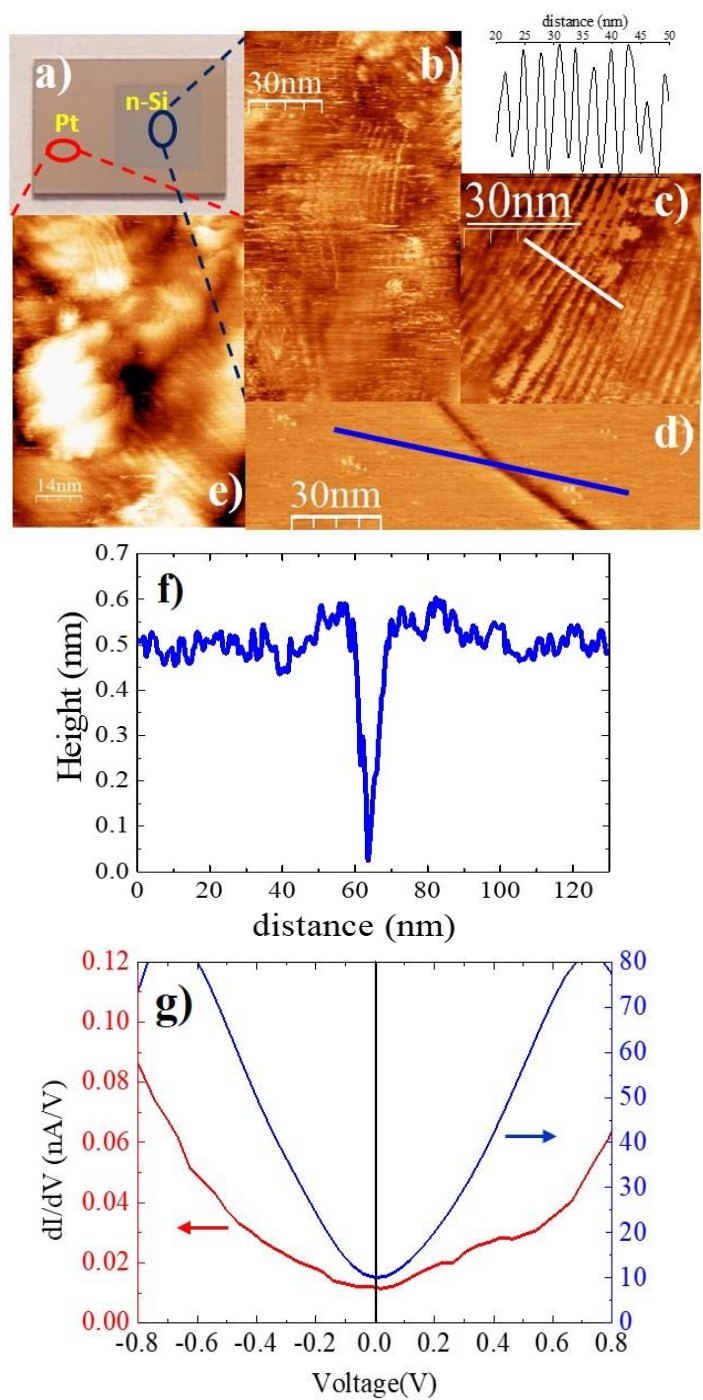

**Fig. 5** STM analysis of a 9.6 nm thick $Bi_2Se_3$ film deposited on Si-Pt substrate. a) Optical image of a $Bi_2Se_3$ film deposited on Si-Pt substrate with the indication of the analysed zones. b) STM of the $Bi_2Se_3$ film deposited on the n-Si side with striped and flat zones. c) Detail of a striped area with the line profile reported on top indicating the stripe periodicity. d) Detail of a flat area with two grains separated by a boundary. The line profile is given by the blue data in f). e) Same as b) but acquired on the Pt part of the same substrate. f) line profile along the blue line in d). g) Average differential conductivity measured by STS on a 150 nm x 150 nm area for a 9.6 nm (blue line) and a 3.7 nm (red line) thick samples. $V$=0 corresponds to the Fermi level.

**boundary.** Fig. 5e shows the STM measurements performed on the $Bi_2Se_3$ film deposited on the Pt side of the same substrate. Here the surface is not flat on a large area as in the case of Si substrate, but a number of islands are present probably due to the substrate morphology. This consists, in fact, of Pt layers 150 nm thick deposited on $SiO_2$ buffered Si wafers. Although Pt grows (111) oriented as shown by XRD measurements, islands growth is quite common in thermal deposited films especially when the substrate is not heated during the deposition as in the present case. Nevertheless, the local $Bi_2Se_3$ surface morphology appears very similar to that observed on Si substrate with the stripes evidenced on each grain, confirming once again the weak influence of the substrate on the lattice structure in this kind of growth.

The local electronic properties of the outermost film surface have been investigated by STS by using the same STM apparatus **soon after having collected the same STS on clean HOPG spectrum. The tunnelling current was registered as a function of the applied bias (*I-V*), the feedback loop was disabled, and the set-point current, which regulated the tip-sample distance, remained unchanged during the voltage scan. The *I-V* curves were collected over grids of points equally spaced on the scanned sample area, and the *I-V* spectra were averaged over a set of several curves.** The blue and red curves in Fig. 5f show the average conductance (obtained by the first derivative *dI/dV* of the current-voltage characteristics) vs. bias voltage acquired on a 150 nm x 150 nm large area for a 9.6 nm and a 3.7 nm thick samples, respectively. **The average was performed on *I-V* characteristics acquired every 1.5 nm. Despite the very local character of the STS analysis, no significative differences were evidenced moving throughout the whole analysed grid of points of the same sample.** The non-zero value of the conductance at zero voltage (which corresponds to the energy position of the Fermi level) confirms the expected metallic character of the $Bi_2Se_3$ thin film surface. This result confirms the success of the deposition technique in obtaining topological insulators with a conductive surface. Moreover, the sizeable decrease of the conductance observed for the thinner sample (see different axis scales in the figure) indicates that the metallic character of the film surface strongly depends on the thickness. A very remarkable feature of this measurement is the position of the minimum in the differential conductance curves. The blue curve in the figure shows the characteristic V shape, smeared by the room temperature effect, representing the Dirac dispersion.[26,27] The minimum energy of the blue curve represents the Fermi energy level; since it is positioned at *V*=0, it indicates that, for these samples, the Fermi level lies at the Dirac point. Although this should be expected for all the TIs, this is the first time, to our knowledge, that it is experimentally observed in the case of $Bi_2Se_3$ samples measured at room temperature and without any treatment of the surface. The Fermi level is usually observed inside the conduction band and this shift is well ascribed to an n-doping of the $Bi_2Se_3$ structure caused by Se vacancies.[2,15,16] In our case instead, the position of the Fermi level at the Dirac point is interpreted as due to the quasi-perfect stoichiometry achieved for these samples because of the Se addiction during the second procedure of the deposition process, as confirmed by XPS measurements. The characteristic V shape is further smeared for thinner samples (red curve in Fig. 5f) as expected.[16]

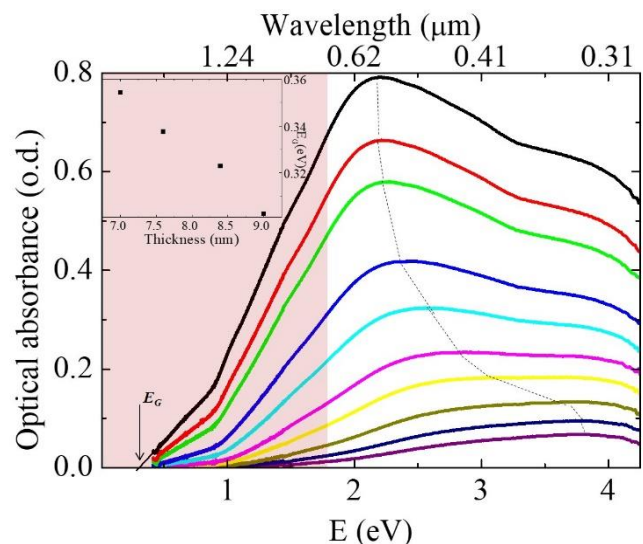

**Fig. 6** Optical absorbance for $Bi_2Se_3$ film with different thickness (from the top: 9.0, 8.4, 7.6, 7.0, 6.4, 5.8, 5.1, 4.5, 3.8, 3.2 nm) deposited on a glass substrate. The black dashed line is a guide to the eyes for the main peak position shift. The light red shadowed area indicates the IR part of the spectrum. The straight line is a linear extrapolation to zero absorption for bandgap estimation in the case of the 9.0 nm thick sample. Inset: bulk gap energy vs. thickness as obtained by the extrapolation of the experimental data reported in the main panel.

## 3. Electro-optical devices

The optical absorbance of the $Bi_2Se_3$ films deposited on glass is measured in the wavelength range $\lambda$=290-3000 nm (*E*=4.25-0.41 eV) as a function of their thickness and reported in Fig. 6 after subtracting the substrate signal. The spectra show a main peak in the visible range, whose position (indicated by the dashed line) changes from *E*=2.18 eV to *E*=3.80 eV with decreasing the film thickness. The peak at *E*=2.18 eV ($\lambda$=570 nm) for the 9.0 nm thick film, can be considered as the 2.1 eV absorption peak commonly observed in $Bi_2Se_3$ bulk samples.[28] Its origin is well documented and it is ascribed to interband transitions between the valence and conduction band energy levels of bulk $Bi_2Se_3$ in different points (F, $\Gamma$, L and Z) of the Brillouin zone.[29] Decreasing the film thickness, the main peak decreases and shifts toward higher energy giving an almost flat spectrum for ultra-thin films resembling features observed in metallic thin layers.[30] This can be interpreted as a rearrangement of the electronic energy levels due to the loss of the bulk properties of the $Bi_2Se_3$ ultra-thin films.[29] Although the main absorbance is in the visible range, a considerable quantity of light is also absorbed in the IR part of the spectrum, indicated by the shadowed area in the figure. As expected, the absorbance increases in the IR range increasing the film thickness giving considerable values for thickness greater than 5 nm and confirming the possible use of $Bi_2Se_3$ as IR photodetectors. By extrapolating the data to the zero absorbance condition it is possible to obtain a rough estimate of the bulk bandgap $E_G$ of the $Bi_2Se_3$ layers, which is reported in the inset of the same figure as a function of the TI thickness. The obtained values, calculated only for the thickest films, show the decrease of $E_G$ with the increase of the film thickness as expected for $Bi_2Se_3$ very thin films.[31] These values are in the

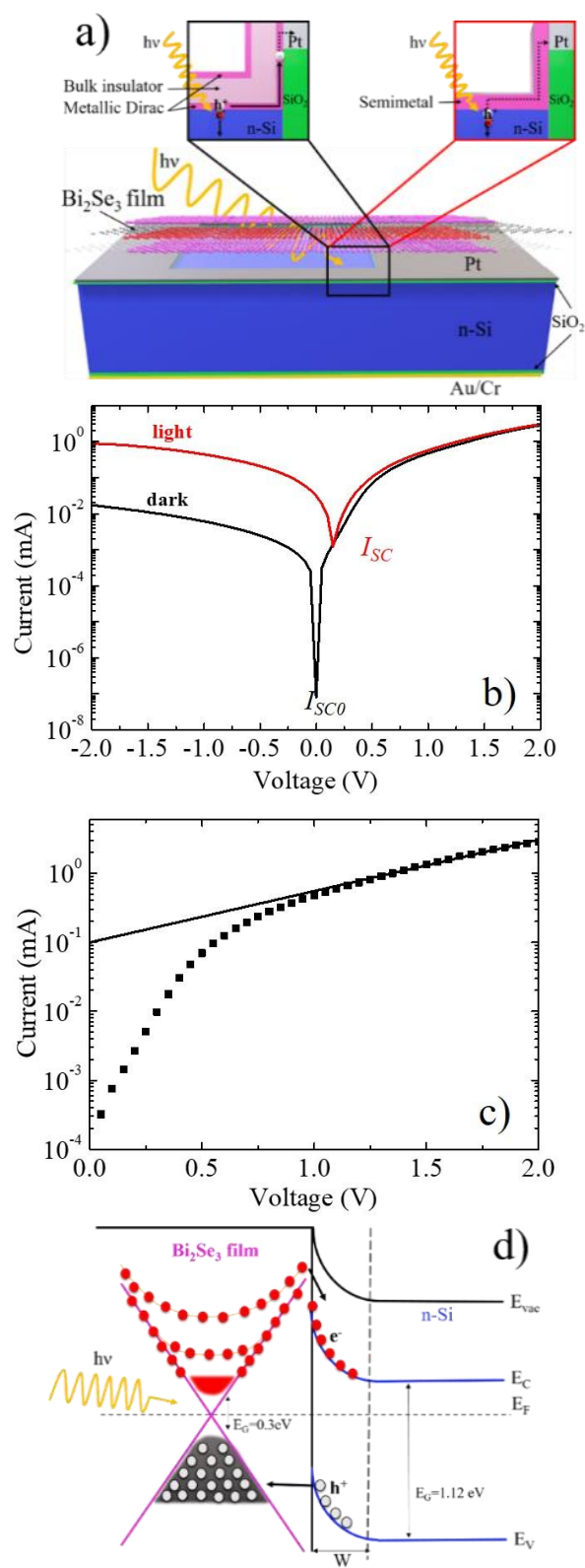

**Fig. 7** a) Schematic view of the Si-Pt substrates used for photodevices. The top left and right sketches are a particular of the $Bi_2Se_3$/n-Si junction in proximity of the Pt electrodes for a thick and a thin $Bi_2Se_3$ film, respectively. b) Current-Voltage characteristics of $Bi_2Se_3$/n-Si heterojunction in dark conditions (black line) and under illumination (red line). c) positive biased I-V characteristic in dark condition shown in b) with the linear fit. d) Sketch of the electronic band bending across the $Bi_2Se_3$/n-Si heterojunction.

range 0.30-0.36 eV in very good agreement with the value of 0.32 eV accepted for this material in bulk form.[6]

The possibility to use the present deposition method for producing optoelectronic devices was explored depositing $Bi_2Se_3$ on the Si-Pt substrates[32] already adopted for graphene[33] and carbon nanotubes/n-Si photodevices.[34] A more detailed sketch of the devices is shown in Fig. 7a. These consist of n-doped Si(00l) wafers 7 mm x 5 mm wide with the top surface partially covered with a Cr/Pt electrode thermally deposited on 300 nm thick $SiO_2$ template layer. The bottom surface of the Si is completely covered with a 150 nm thick Cr/Au metallic layer forming an Ohmic contact with Si. When a film is deposited on the top surface, a rectified junction is formed with the n-Si part, with lateral size of 3 mm x 3 mm, not covered by $SiO_2$/Cr/Pt, which plays the role of active area of the photosensitive devices. In this case, the top and the bottom metallic contacts act as electrical terminals for the collection of the photogenerated charges.

The success of our deposition method in fabricating $Bi_2Se_3$/n-Si heterojunctions is confirmed by the current-voltage (I-V) characteristics shown in Fig. 7b recorded in dark conditions (black line) for a 9.6 nm thick $Bi_2Se_3$ layer deposited on the Si-Pt substrate. The curve shows good rectified properties with an on-off current ratio of more than 2 order of magnitudes at $V=\pm 2.0$ V. Similar results, but with a smaller value of the on-off ratio, are obtained for thinner $Bi_2Se_3$ layers. The dark current ranges between 0.2 $mA \cdot cm^{-2}$ and $1 \times 10^{-6}$ $mA \cdot cm^{-2}$ for reverse voltage in the range from -2.0 V to 0 V. These values are much lower than that observed by other authors for similar junctions[7] and comparable to topological crystalline insulators such as SeTe/Si.[35] The performance of the junctions can be limited by the granularity of the $Bi_2Se_3$ film, which reflects on its electrical resistance due to the presence of grain boundaries. The slope of the I-V data in the range of high positive voltage gives an estimation of this resistance. Fig. 7c shows a fit to the data, acquired in dark conditions, in this limited range. The linear fit gives a resistance value of 1.35 k$\Omega$, roughly corresponding to a resistivity of $5.7 \times 10^{-4}$ $\Omega \cdot cm$ in very good agreement with $Bi_2Se_3$ obtained by other techniques.[5] This result suggests that the granularity does not consistently reduce the sample quality.

The photovoltaic properties of the junctions are investigated by illuminating the active area with a solar simulator (AM1.5) 100 $mW \cdot cm^{-2}$ white light. The sketch in Fig. 7d is representative of the band bending at the $Bi_2Se_3$/n-Si interface. After light absorption, the photogenerated charges are drifted across the junction by the interface potential and collected by the electrodes. The corresponding I-V characteristic of the junction under illumination is reported in Fig. 7b (red curve). The effect of the light is to improve the short circuit current ($I_{SC}$) of more than 4 order of magnitudes with respect to the dark condition. Fig. 8a shows the responsivity $\mathcal{R}$, for all the obtained samples, given by the expression $\mathcal{R} = (I_{SC} - I_{SC0})/P$ where $I_{SC0}$ is the short circuit current in dark condition and $P$ is the power of the incident light. Responsivity up to 1 $A \cdot W^{-1}$ is obtained for the junctions with the thickest $Bi_2Se_3$ layers whereas a strong reduction of the detected light is observed for thickness below 7 nm. Interestingly, for a given thickness, the responsivity also depends on the reverse bias voltage showing a consistent increase in the performance of the detector for voltages as low as -10 V. Assuming a dark current $I_{SC0}$=10 nA for all the obtained devices, a Noise Equivalent Power $\mathcal{NEP} = I_{SC0}/\mathcal{R} \sim 10^{-11}$ W·Hz$^{-1/2}$ is obtained.

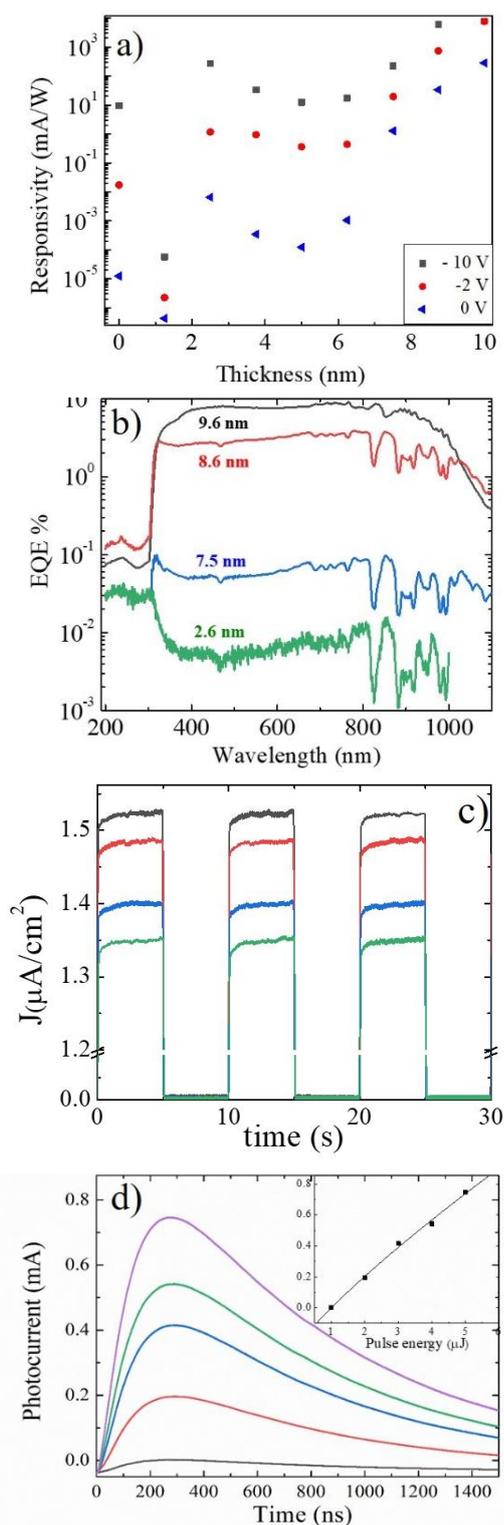

**Fig. 8**. (a) Responsivity vs. thickness for $Bi_2Se_3$/n-Si heterojunctions at different reverse bias voltages. (b) EQE vs. wavelength for $Bi_2Se_3$/n-Si heterojunction with different thickness of the $Bi_2Se_3$ film. (c) Photocurrent density vs. time at different wavelength (from the bottom $\lambda$=460 (green), 568 (blue), 595 (red), 660 nm (black)) under on-off light excitation of the 9.0 nm thick sample. (d) Time response of the same sample to a femtosecond laser pulse at $\lambda$=1500 nm for different pulse energy (from bottom: 1, 2, 3, 4, 5 µJ); inset: maximum of the photocurrent vs. laser pulse energy: the line is fit to the data using the power law $y=x^{0.84}$.

The ability of a photodetector to detect small signals is given by the detectivity $\mathcal{D} = \sqrt{A/2eI_{SC0}}\,\mathcal{R}$ where $A$=9 mm$^2$ is the active area, $I_{sc0}$=10$^{-8}$ A and $e$ is the elementary charge. Considering the thickest sample, values of $\mathcal{D}$ in the range 10$^{10}$-10$^{12}$ Jones are obtained for bias voltages up to -10 V. These results encourage the possible use of these junctions as photomultiplier detectors operating in low voltage regime. The thickness dependence of the device performances is confirmed by the data shown in Fig. 8b where the external quantum efficiency (*EQE*) is reported as a function of the wavelength. *EQE* gives the number of produced photocharges for an incident photon and it is defined by the expression $EQE = hcI_{SC}/(q\lambda P)$ where $h$ is the Planck constant, $c$ the speed of light, $q$ the elementary charge. In agreement with the results shown in Fig. 6, *EQE* increases with the $Bi_2Se_3$ thickness due to the increase in its optical absorbance. Nevertheless, for the thinnest layer (2.6 nm in the figure) a very negligible quantity of photocharges is produced.

Fig. 8c shows the photodetector response to an on-off light excitation obtained by using a light power of 0.25 mW·cm$^{-2}$ emitted by a LED operating at different wavelengths. The data, recorded at *V*=0, are a proof of the excellent repeatability of the device response which is independent on the wavelength and it remains of the same value after hundreds of switching cycles. The ability of the photodevices to detect IR light was investigated by illuminating the active area by a femtosecond laser pulse at $\lambda$=1500 nm and detecting the photocurrent by using a 50 GHz oscilloscope.[33] Fig. 8d reports the photocurrent at *V*=0 produced by the device, for different energy of the incident beam. The response is remarkable and a measure of the rise time (between 10% and 90% of the full current peaks) gives 130 ns for all the reported curves, which is much lower than data reported for previous TI ultrafast photodetectors.[36] The inset of the same figure shows the maximum photocurrent, measured at the maximum of each curve in the main panel, as a function of the pulse energy. The line is a fit to the data using the power law $y=x^\vartheta$ obtained with $\vartheta$=0.84 which indicates the good linearity of the photodetector in the IR range.

## 4. Discussion

The V shaped differential conductance centred at *V*=0, reported in Fig. 5f, clearly indicates that, at least locally, the Femi level lies in the centre of the energy bandgap as expected for $Bi_2Se_3$ with ideal composition. This important result, not observed experimentally before because of the lack of Se in this material, is observed here thanks to the deposition method, which allows achieving the correct film stoichiometry by adding Se during the second procedure of the fabrication process.

The position of the Fermi level at the centre of the bandgap affects the photodetector performances and influences the choice of the counter electrode material. In fact, when, as in the common cases, the Fermi level is shifted towards the conduction band, $Bi_2Se_3$ film shows n-doping behaviour. In this case, a p-doped counter electrode must be chosen in order to maximize the on-off current ratio. With the Fermi level at the centre of the bandgap, no matter whether p-doped or n-doped counter electrode is used giving a much high flexibility in the heterojunction fabrication process.

The experimental data also show that our method is suitable for ultra-thin film growth on many kinds of substrates (both solid and flexible). In particular, in the case of n-doped Si substrates, it allows to obtain heterojunctions for photovoltaic applications in a very easy and fast way. However, the strong reduction of the photodetector performance for $Bi_2Se_3$ thickness less than 7 nm (reported in Fig. 8a and 8b) deserves some discussion.

Fig. 7c shows an electronic band sketch of the $Bi_2Se_3$/n-Si heterojunction that, although schematically, helps to understand the photocharge generation. When the thickness of the $Bi_2Se_3$ film is greater than 7-8 nm, the incident photons are in part absorbed by the $Bi_2Se_3$ layer and in part absorbed by the underlying n-Si substrate at the active window. The produced photocharges are then separated by the barrier potential at the $Bi_2Se_3$/n-Si interface and delivered to the Au bottom and Pt top contacts.[32-34] In particular, the photocharges reach the Pt electrode travelling as a 2D Dirac electron gas through the $Bi_2Se_3$ top layer thanks to its surface properties (see sketch in Fig. 7a, top left). When the thickness of the $Bi_2Se_3$ layer is below 5-6 nm, the photoresponse is strongly reduced even though the rectified properties of the junctions are still achieved. This is consistent with the general properties of $Bi_2Se_3$ where the Dirac metallic states are formed only for thickness greater than 6 QLs.[16] Thinner layers lose the Dirac point due to the interference between the electron wavefunctions of the two opposite surfaces. This gives rise to the formation of a small surface bandgap, which replaces the metallic Dirac cone with the consequence of losing part of the 2-D electron gas character of the surface. This behaviour is confirmed, for our samples, by the STS measurements reported in Fig. 5f where a reduction of the surface conductance is measured by reducing the film thickness. The photogenerated charges are then constrained in a layer with higher electrical resistance during their travelling towards the Pt electrodes (see sketch in Fig. 6a, top right). Therefore, the reduction of the thickness has a double detrimental effect on the photodetector performance: i) a reduced number of photocharges is produced inside the $Bi_2Se_3$ layer and ii) the photocharges produced in the n-Si substrate reach the Pt electrode through the $Bi_2Se_3$ layer which has a gap at the Dirac point with a consequent increase of its electrical resistance. This behaviour also explains the decrease of $\mathcal{R}$ for thickness up to 6 QLs. In this thickness range, $Bi_2Se_3$ film plays the role of optical window whose transparency decreases with the thickness. This reduces the number of photons that reaches the interface and, as a consequence, the number of generated photocharges and $\mathcal{R}$.

Photovoltaic response was already investigated for $Bi_2Se_3$/Si heterojunctions.[4,7,12,37-40] The best results were obtained by using $Bi_2Se_3$ nanoflakes and Si nanowires where[40] $\mathcal{R}$~$10^3$ A·W$^{-1}$ and $\mathcal{D}$~$10^{12}$ Jones.[7,12] Other photodevices, based on $Bi_2Se_3$ nanowires/doped-Si substrates, showed $\mathcal{R}$ as high as 300 A·W$^{-1}$.[38] These high values are due to the very small active area illuminated, which imply a small junction area with uniform interface properties. Moreover, on-off ratio of the order of 10 are measured by I-V characteristics.[38] In comparison, our experimental results give a lower value of the responsivity but high on-off current ratio. The low responsivity can be related to the presence of grain boundaries in the $Bi_2Se_3$ layer, remarkable in very thin films, which act as traps that enhance the photocharge recombination mechanism. On the other side, a high on-off current ratio indicates a good junction quality.

An improvement in the photovoltaic properties can be achieved by increasing the $Bi_2Se_3$ thickness. This can be regulated by moving the substrates in a different position during the second procedure of the fabrication process or by increasing the amount of $Bi_2Se_3$ in the quartz boat before starting the whole process. Another way to deposit thicker films is to reduce the pumping speed during the outgassing of the species from the inner wall of the tube. In this way, the species spend more time in the substrate zone during the condensation stage with the result to obtain thicker layers.

## Conclusions

A new deposition method for $Bi_2Se_3$ ultra-thin films has been demonstrated using the by-products of a previous nanowires/nanobelts growth. The method allows obtaining many samples (up to 10) with different thickness and on different substrates with the expected stoichiometry. This has allowed obtaining for the first time TIs with the Fermi level correctly positioned in the middle of the bulk bandgap, a property achieved by other methods as surface cleaving in bulk samples under UHV conditions. The moderate temperature reached during the growth (less than 100 °C) gives to this method the access to a large kind of substrates such as multilayers, metallic contacted and low temperature melting materials as plastic tapes. The films have been successfully used to realize wideband photosensitive $Bi_2Se_3$/Si heterojunctions with good electro-optical properties suggesting a new and cheap method for IR photodetector production. Since the growth mechanism is based on the van der Waals structure of $Bi_2Se_3$, it can be applied to other materials having the same specific properties as TIs and transition metal dichalcogenides.

## Conflicts of interest

There are no conflicts to declare.

## Acknowledgements

A special thank goes to Dr. J. Andzane for important suggestions on the growth process nanowires/nanobelts. Thanks to C. D'Ottavi for XRD measurements. M. Salvato, M. Scagliotti, P. Castrucci and M. De Crescenzi would like to acknowledge the European Community for the Horizon 2020 MSC-RISE Project DisetCom (GA823728). M. Salvato also would like to thank the COST Action CA16218 Nanoscale Coherent Hybrid Devices for Superconducting Quantum Technologies. The work has been also supported by the Horizon 2020 research and innovation program (Grant Agreement No. 766714/HiTIMe).